\documentclass[referee,useAMS,usenatbib]{mn2e}   
\usepackage{amssymb}
\usepackage{graphics}
\usepackage{lscape}
\usepackage{amssymb}
\def\etal{\textit{et al.}\ }
\def\ie{i.\,e.\,, }
\def\eg{e.\,g.\,}

\def\near{$\sim$}
\def\pmi{$\pm$}

\def\Ha{H$\alpha$\ }

\title[Kinematics of the YSOs and the CGs in Gum Nebula]
{Kinematics of the Young Stellar Objects associated with the   Cometary   
Globules in  the Gum Nebula }

\author[R. Choudhury  et al.]  {Rumpa Choudhury$^{1,2}$\thanks{E-mail:
rumpa@iiap.res.in},  H. C.  Bhatt$^{1}$\\  $^{1}$ Indian  Institute of
Astrophysics, Bangalore 560034,  India\\ $^{2}$ University of Calicut,
Calicut 673 635, India}

\begin{document}
\date{}
\pagerange{\pageref{firstpage}--\pageref{lastpage}} \pubyear{2008}
\maketitle
\label{firstpage}


\begin{abstract}

An analysis of proper motion measurements of the Young Stellar Objects
(YSOs) associated with  the Cometary Globules (CGs) in  the Gum Nebula
is  presented.  While  earlier studies  based on  the  radial velocity
measurements of the CGs suggested  expansion of the system of the CGs,
the  observed  proper  motion  of  the  YSOs  shows  no  evidence  for
expansion.  In particular  the kinematics of two YSOs  embedded in CGs
is  inconsistent with  the  supernova explosion  of  the companion  of
$\zeta$ Pup about 1.5 Myr ago as  the cause of the expansion of the CG
system.  YSOs associated with the  CGs share the average proper motion
of the member stars of the  Vela OB2 association. A few YSOs that have
relatively  large proper  motions  are found  to  show relatively  low
infrared excesses.

\end{abstract}



\section{Introduction}

The  Gum Nebula, a  prominent HII  region of  southern Milky  Way, was
discovered by  (Gum 1952).  The  whole nebula is around  36$^\circ$ in
diameter (Chanot \& Sivan 1983). The  true nature of the Gum Nebula is
not  very clear.   Along with  the two  O type  stars $\zeta$  Pup and
$\gamma^{2}$ Velorum,  two OB associations,  Vela OB2 and Tr  10, have
been  found in the  Gum Nebula  region. There  are also  two supernova
remnants (SNRs)  in the  direction of  the Gum Nebula,  Pup A  and Vela
SNR. However, Sridharan (1992)  argued that the spatial coincidence of
these two  SNRs with  the Gum Nebula  is due to  chance superposition.
Chanot \&  Sivan (1983)  described the Gum  Nebula as  an intermediate
structure between classical HII region and typical supernova remnant.

About 30 cometary globules have been  found in the Gum Nebula and they
are  distributed in  a  nearly circular  pattern.   They share  common
features of a compact, dusty head (sometimes with a bright-rim), and a
comet-like tail.   The tails of the globules  generally point radially
outwards from  the centre  of the Vela  OB2 association.  The  axes of
these  CGs  seem   to  converge  on  a  very   small  area  around  at
l$_{\parallel}$\hspace{1mm}                           \near\hspace{1mm}
261$^\circ$,          \hspace{1mm}         b$_{\parallel}$\hspace{1mm}
\near\hspace{1mm} -\hspace{1mm} 5$^\circ$  \eg  Reipurth (1983), Zealey
et al., (1983).  The source of the CG complex is as yet still unclear.
A  single source  may not  be responsible  for the  formation  and the
evolution of  the Gum Nebula  complex.  $\zeta$ Pup,  and $\gamma^{2}$
Velorum and  the two  OB associations are  considered as  the probable
source of  the UV radiation and  photoionisation of the  nebula in the
literature.

There are  some uncertainties about  the distances of $\zeta$  Pup and
$\gamma^{2}$  Velorum.   The   estimated  Hipparcos  distance  to  the
Wolf-Rayet   WC8+O  spectroscopic   binary  $\gamma^{2}$   Velorum  is
258$^{-31}_{+41}$  pc  and to  the  O4I(n)f  star  $\zeta$ Pup  it  is
429$^{-77}_{+120}$  pc (van  der  Hucht \etal  1997).  However,  Pozzo
\etal (2000) argued that the  distance to the $\gamma^{2}$ Velorum may
be same  as that  for the Vela  OB2 association.  Again  the estimated
Hipparcos distance  to the  Vela OB2 association  is 410 pc  (de Zeeuw
\etal 1999) and  Woermann \etal (2001) estimated a  distance of 500 pc
to the expansion centre of the association. In this context an average
distance  to the  CG system  as  450 pc  is reasonable  as adopted  by
Sridharan (1992).

There have  been several studies on  the kinematics of  the Gum Nebula
region  to determine  whether  the  system is  expanding  or not,  \eg
Yamaguchi \etal  (1999) and  references therein.   Most of  the studies
concluded that  the molecular material associated with  the Gum Nebula
is  expanding though  the expansion  velocities obtained  from various
studies  are  different  from  each  other. Zealey  \etal  (1983)  and
Sridharan  (1992) studied  the kinematics  of the  CGs in  details and
based on the radial velocity measurements concluded that the CGs are
expanding and obtained  the expansion velocity of the  system as 5 and
12  km s$^{-1}$  respectively. Woerman  \etal (2001)  investigated the
kinematics  of  the neutral  material  around  the  Gum Nebula.   They
concluded  that  the   Diffuse  Molecular  Clouds (DMCs)  and  Cometary
Globules(CGs) form  a single  expanding shell centred  on \hspace{1mm}
l$_{\parallel}$\hspace{1mm}=\hspace{1mm}       261$^\circ$,\hspace{3mm}
b$_{\parallel}$  =\hspace{1mm} -2.5$^\circ$. According to  their model
the shell is asymmetric with the  radii of the front and back faces as
130  and  70  pc  respectively.   They  also  obtained  the  expansion
velocities  of  14  and  8.5  km $s^{-1}$  for  front  and  back  faces
respectively. They suggested the  supernova explosion of the companion
of $\zeta$  Pup about 1.5  Myr ago as  the probable origin of  the Gum
Nebula and the expanding shell.

There have been no studies of the transverse (in the plane of the sky)
motion of  the Gum Nebula  and the associated  CGs.  In this  paper we
have  examined the  proper motion  measurements of  stellar  and young
stellar objects  (YSOs) in this region  to study the  expansion of the
system of CGs and possible  sources responsible for the triggered star
formation in the CGs of Gum  Nebula.  In Section 2. and 3. we summarise
the characteristics of the CGs in  the Gum Nebula as well as the known
YSOs in and around the CGs.  We discuss the proper motions of the YSOs
in Section 4.  Results are  discussed in Section 5. and conclusions are
presented in Section 6.

\section{Cometary Globules in the Gum Nebula}

\begin{figure*}
\resizebox{16.4cm}{16cm}{\includegraphics{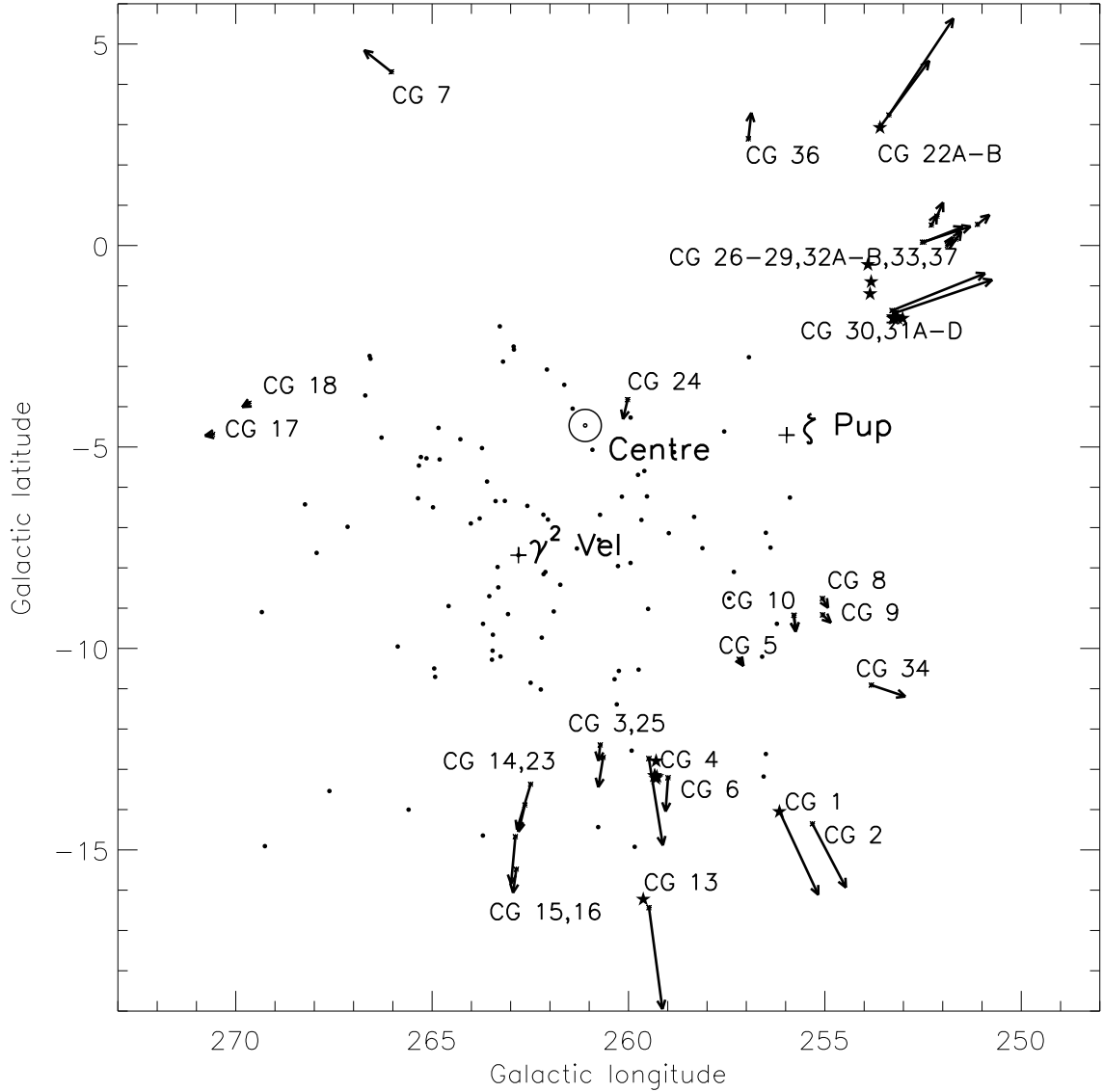}}
\caption{Distribution  of CGs  in  the galactic  coordinate. The  tail
  vectors  have  lengths  proportional  to  the tail  lengths  of  the
  respective  CGs.  \textit{Filled Circles} represent  the members  of the  Vela OB2
  association. \textit{Stars}  represent the  Young Stellar Objects  around the
  CGs.  Positions of  $\zeta$  Pup and  $\gamma^{2}$  Velorum are  also
  marked .}
\label{cg-l-b.ps}
\end{figure*}

Zealey \etal  (1983) carried  out an extensive  study of  the cometary
globules in the Gum Nebula region.   They found 29 CGs within a region
of  projected  angular  radius  of 9.5$^\circ$.   They  tabulated  the
coordinates, position  angles, tail directions  and measured H$_{2}$CO
radial velocities  of some of  the CGs. They proposed  the approximate
centre of  the CG complex based on  the best fitting circle  of the CG
positions.    Sridharan  (1992)  found   some  discrepancies   in  the
coordinates of the CGs and redefined the coordinates of the individual
CGs. Sridharan (1992) studied  the kinematics of the cometary globules
in  the  Gum  Nebula  using  the transition  lines  of  $^{12}CO$  and
tabulated  the V$_{LSR}$  velocities  of  most of  the  CGs.  We  have
adopted  positions  and  velocities  from  Sridharan  (1992)  and  the
approximate centre  of the system from Zealey  \etal (1983).  Reipurth
(1983) suggested that the CGs  are pointed towards the triangle formed
by $\zeta$ Puppis, $\gamma^{2}$  Velorum and Vela Pulsar and estimated
an average projected distance of 70  pc for the CGs from the centre of
the  triangle.  It  is possible  that  the two  O type  stars and  the
progenitor of the Vela Pulsar were the main energy source of the whole
Gum  Nebula region  in the  past.  In  Figure \ref{cg-l-b.ps}  we have
plotted the  respective CGs, members  of the Vela OB2  association and
the  Young  Stellar  Objects(YSOs)  around  the CGs  in  the  galactic
coordinates together with the the probable energy sources $\gamma^{2}$
Velorum and $\zeta$ Pup. The  adopted centre of the nebula from Zealey
\etal (1983) is also plotted.   The vectors associated with the CGs in
the plot have lengths proportional  to the tail-lengths of the CGs and
also indicate their directions.

Various authors  derived the  centre of expansion  of the  CG complex,
\eg,  Zealey \etal (1983),  Reipurth (1983)  and Sridharan  (1992). In
this paper we have adopted the  position of the centre as suggested by
Zealey                 \etal                 (1983)                \ie
RA(2000)\hspace{1mm}=\hspace{1mm}08$^h$19$^m$40.54$^s$\hspace{3mm}
Dec(2000)\hspace{1mm}=\hspace{1mm}-\hspace{1mm}44$^\circ$09$'$30.8$''$
(l\hspace{1mm}\near\hspace{1mm}261.10$^\circ$,\hspace{3mm}
b\hspace{1mm}\near\hspace{1mm}    -\hspace{1mm}4.46$^\circ$)\hspace{3mm}
which   is   in   good    agreement   with   the   centre   coordinate
(l\hspace{1mm}\near     \hspace{1mm}     261.5$^\circ$    ,\hspace{3mm}
b\hspace{1mm}\near\hspace{1mm}-\hspace{1mm}4.0$^\circ$)\hspace{3mm}
obtained  by   Woerman  \etal  (2001).   We  have   also  adopted  the
approximate    centre    of    the    Vela    OB2    association    as
RA(2000)\hspace{1mm}=\hspace{1mm}08$^h$07$^m$00.22$^s$\hspace{3mm}
Dec(2000)\hspace{1mm}=\hspace{1mm}-\hspace{1mm}46$^\circ$41$'$53.8$''$
(l\hspace{1mm}\near\hspace{1mm}262.3$^\circ$,\hspace{3mm}
b\hspace{1mm}\near\hspace{1mm}-\hspace{1mm}7.71$^\circ$).

Reipurth (1983)  proposed a mechanism for the  formation and evolution
of the  CGs based on  the interaction of  the massive stars  and their
parent molecular cloud.  Brand \etal (1983) proposed that the cometary
morphology has  been created  by the passing  of shock wave  through a
spherical molecular cloud. They also suggested  the possibility of
star  formation in  the shocked  molecular clouds.   Analytical models
(also known as the radiation driven implosion (RDI) models ) have been
developed by  Bertoldi (1989) and Bertoldi \&  McKee (1990).  Lefloch \&
Lazareff (1994)  studied the evolution and morphology  of the cometary
globules  by a  2-d hydrodynamic  simulation. These  models  could not
reproduce the details of triggered  star formation by the UV radiation
because they  did not include the  effect of self  gravity. Miao \etal
(2006) have  shown the evolution of  the cometary globules  as well as
the triggered star  formation under the influence of  the UV radiation
of  the massive stars  by a  3-d smoothed  particle hydrodynamics (SPH)
simulation.

According  to  these  models,  when  an  OB  association  forms  in  a
comparatively dense core  of a clumpy Giant Molecular  Cloud (GMC) its
interaction  with its  parent molecular  cloud can  be  destructive or
constructive  depending on the  size and  density distribution  of the
leftover clumps  and cores of the  parent GMC.  The  UV radiation from
the massive young stars evaporates  the gas and less dense material in
its surroundings  and creates an  expanding Str\"{o}mgren sphere.   But it
cannot evaporate the relatively dense cores of the clumpy parent cloud
because of the recombination shielding of the front layer known as the
ionised boundary  layer [IBL]  (Elmegreen 1976). However,  the ionised
gas starts  to move  towards the ionising  source and due  to momentum
conservation the clumps also get  a velocity radially outward from the
central star.   This mechanism  is known as  Rocket Effect  (Oort $\&$
Spitzer  1950).  The rocket  effect pushes  the loosely  bound gaseous
envelope  much more effectively  than the  dense core.   The expanding
Str\"{o}mgren sphere  also creates a  shock wave which  interacts with
the  leftover  remnant  clumps  and  converts  them  into  more  dense
regions. As a  result a newly exposed dense core  generally has a thin
ionized  layer and a  tail in  the opposite  direction created  by the
leftover material of the parent GMC.  At the same time the compression
by the shock  front can trigger the next  generation star formation in
the  dense core.  It thus  provides the  additional  external pressure
force  to the  gravity to  collapse the  clumps to  form  stars.  This
process  of star  formation is  known as  triggered star  formation as
opposed  to  the spontaneous  star  formation  where  only gravity  is
responsible for  the collapse  of the cloud  core.  Also the  low mass
stars  which have  formed in  these globules  simultaneously  with the
massive stars are exposed to photoevaporation.

Considerable  evidence for  current  low mass  star  formation in  the
cometary  globules of  the Gum  Nebula is  discussed in  the following
section.

\section{  Young Stellar Objects in and around  the Cometray Globules}

Photometric and spectroscopic studies to investigate the low mass star
formation   in  the   Cometary  Globules   have  been   undertaken  by
Petterson (1987), Sahu (1992), Reipurth  (1983, 1993), Kim (2005).  Based
on H$\alpha$  emission line, Li  absorption line and near  IR infrared
excesses,  \near\hspace{1mm}  30  YSOs  have been  identified  in  the
direction of the Gum Nebula. There are confirmed signatures of the low
mass star formation  in the globules; \eg CG  1(Reipurth, 1983), CG 22
[Sahu  (1992), Reipurth (1983,  1993)], CG  30/31/38  complex [Reipurth 
(1983),  Petterson (1987),  Kim (2005)],  CG  4/CG 6/Sa101  and CG  13
(Reipurth,  1993).    We  have  tabulate  the  optical   and  near  IR
photometric  measurements   of  the  individual  YSOs   in  the  Table
\ref{spec_phot_data}.   We also  include  the spectral  types and  the
measured     H$\alpha$    equivalent     widths    in     the    Table
\ref{spec_phot_data}.   Near IR  magnitudes are  taken from  the 2MASS
Point  Source Catalog (Cutri,  2003). Except  for NX  Pup which  is of
spectral type A,  the other YSOs found near the CGs  are all late type
stars with  spectral types  F, K and M.   We also tabulate  some useful
parameters related  to the known star  forming CGs and  dark clouds in
Table \ref{cloud_param}.

Petterson   (1987)  made   photometric,  spectroscopic   and  infrared
observations of  16 H$\alpha$ emission  line stars in the  region near
the cometary globules  CG 30/31/38.  They identified 9  T Tauri stars.
They  also   found  that   except  PHa  12   and  PHa  44   all  other
spectroscopically confirmed YSOs  are variable in V band.   PHa 41 has
shown optical variability of \near\hspace{.8mm} 3  mag. and PHa 15, PHa 21, PHa 34,
PHa 41 have large (U-B) excesses.

Petterson \&  Reipurth (1993) surveyed  five fields in the  Gum Nebula
for low mass emission line stars. They found 7 H$\alpha$ emission line
stars  near the  CG4/CG6/Sa 101  cloud complex  and one  more  near CG
13. They  confirmed the  nature of the  objects as the  low-mass young
stars [RP93] based on the  low resolution spectra, optical and near IR
photometry.

Kim \etal (2005)  presented photometric and spectroscopic observations
of  low mass  pre-main  sequence  stars in  the  cometary globules  CG
30/31/38.  They identified PMS  stars in that direction by photometric
and spectroscopic studies.  They confirmed  the youth of the PMS stars
using the lithium abundances. They also measured the radial velocities
of  the PMS  stars which  are consistent  with those  of  the cometary
globules. However,  Kim \etal  (2005) also suggested  that XRS  9, KWW
1055, KWW 1125,  KWW 1333 and KWW 1806 are probably  old field stars (
50-100 Myr)  with strong magnetic  activity. We do not  consider these
objects in further discussion.

\begin{table*}
\centering
\begin{minipage}{\linewidth}
\caption{Photometric and spectroscopic measurements of YSOs associated with CGs and Diffuse Molecular Clouds}\label{spec_phot_data}
\begin{tabular}{@{}lcccccccc@{}}
\hline
 No.& Name&V        &Spectral   &W$_\lambda$(\Ha)&J     & H     &K         &Ref.$^a$   \\
    &               &         &Type  &A$^\circ$        &      &       &          &   \\

\hline
 1& NX Pup          &10.61$^{(1)}$&A0$^{(1)}$&   -44.0$^{(5)}$    &  8.579 \pmi 0.030 & 7.285\pmi	 0.042 & 6.080 \pmi	 0.031 &1,5 \\
 2& KWW 464	    & 15.82   & M3V  &	 -2.8 	      &  12.126\pmi 0.024 &11.392\pmi	 0.026 &11.173 \pmi	 0.026 &2 \\
 3& KWW 1892/PHa 12 & 15.17   & M1V  &	 -26.6/-31.3  &  11.402\pmi 0.023 &10.663\pmi	 0.023 &10.323 \pmi	 0.021 &2,3 \\
 4& KWW 598	    & 17.27   & M2V  &	 -11.5 	      &  12.150\pmi 0.024 &11.519\pmi	 0.023 &11.281 \pmi	 0.023 &2 \\
 5& KWW 1863        & 14.65   & M1V  &	 -2.8 	      &  10.873\pmi 0.028 &10.194\pmi    0.033 & 9.940 \pmi      0.024 &2 \\
 6& KWW 1637        & 12.15   & K6V  &	 -2.4 	      &   9.529\pmi 0.023 & 8.880\pmi	 0.022 & 8.708 \pmi	 0.024 &2 \\
 7& KWW 873	    & 13.81   & K7V  &	 -7.9 	      &  10.676\pmi 0.023 & 9.947\pmi	 0.023 & 9.578 \pmi	 0.023 &2 \\
 8& KWW 1043/PHa 15 & 16.61   & M3V  &	   --         &  11.925\pmi 0.028 &11.135\pmi	 0.025 &10.628 \pmi	 0.024 &2,3 \\
 9& KWW 975/Pha 14  & 15.56   & M2V  &	 -8.43 	      &  11.387\pmi 0.022 &10.640\pmi	 0.025 &10.299 \pmi	 0.023 &2,3 \\
10& KWW 1302	    & 15.76   & M4V  &	 -8.23 	      &  11.500\pmi 0.032 &10.761\pmi    0.036 &10.429 \pmi      0.023 &2 \\
11& KWW 1953        & 15.58   & M3V  &	 -4.24/ -4.93 &  11.438\pmi 0.023 &10.722\pmi	 0.022 &10.510 \pmi	 0.023 &2 \\
12& KWW 2205        & 16.20   & M4V  &	 -4.34	      &  11.780\pmi 0.023 &11.122\pmi    0.022 &10.840 \pmi      0.021 &2 \\
13& KWW XRS 9       &  --     & G5V  &	  3.23 	      &  9.783 \pmi 0.021 & 9.464\pmi    0.022 & 9.346 \pmi      0.023 &2 \\
14& KWW 1055	    & 14.40   & G2V  &	  2.17 	      &  12.424\pmi 0.026 &12.028\pmi	 0.025 &11.890 \pmi	 0.023 &2 \\
15& KWW 314	    & 15.14   & A3e  &	  6.2         &  12.546\pmi 0.024 &12.155\pmi    0.023 &11.931 \pmi      0.026 &2 \\
16& KWW 1125	    & 12.35   & $<$F8V &  5.75        &  11.285\pmi 0.023 &11.085\pmi	 0.023 &11.015 \pmi	 0.023 &2 \\
17& KWW 1333	    & 13.64   & $<$F8V &  5.23        &  12.074\pmi 0.024 &11.788\pmi	 0.027 &11.694 \pmi	 0.026 &2 \\
18& KWW 1806	    & 13.99   & $<$F8V &  0.81 	      &  12.262\pmi 0.028 &11.820\pmi    0.031 &11.678 \pmi      0.027 &2 \\
19  &PHa 44         & 15.8    &K7-M0 &  -50.7         &  12.996\pmi 0.028 &12.175\pmi	 0.027 &11.713 \pmi	 0.019 &3 \\
20  &PHa 51         & 15.7    &K7-M0 &  -70.1         &  12.664\pmi 0.027 &11.730\pmi	 0.023 &11.090 \pmi	 0.023 &3 \\
 3  &PHa 12         & 15.5    &M1.5  &  -16.1         &  11.402\pmi 0.023 &10.663\pmi	 0.023 &10.323 \pmi	 0.021 &3,2 \\
21  &PHa 21         & 16.4    &M4    &  -48.1         &  12.212\pmi 0.026 &11.420\pmi	 0.022 &11.058 \pmi	 0.023 &3 \\
22  &PHa 34         & 15.6    &K3    &  -60.5         &  12.439\pmi 0.029 &11.642\pmi	 0.026 &11.031 \pmi	 0.023 &3 \\
23  &PHa 40         & 16.4    &M0.5  &  -18.7         &  12.750\pmi 0.024 &11.759\pmi	 0.022 &11.326 \pmi	 0.021 &3 \\
24  &PHa 41         & 14.0    &--    &  -98.6         &  10.775\pmi 0.024 & 9.806\pmi	 0.022 & 8.914 \pmi	 0.024 &3 \\
 8  &PHa 15         & 16.9    &M3    &  -130.5        &  11.925\pmi 0.028 &11.135\pmi	 0.025 &10.628 \pmi	 0.024 &3,2 \\
 9  &PHa 14         & 16.4    &M2    &  -22.0         &  11.387\pmi 0.022 &10.640\pmi	 0.025 &10.299 \pmi	 0.023 &3,2 \\
25  &PHa 92         & 13.38   &K2    &  -35.4         &  10.573\pmi 0.023 & 9.692\pmi	 0.024 & 9.044 \pmi	 0.021 &3 \\
26  &[RP93] 1       & $>$17   &M3-4  &  -24.8         &  11.378\pmi 0.023 &10.716\pmi	 0.025 &10.406 \pmi	 0.023 &4 \\
27  &[RP93] 2       & $>$17   &M2    &  -266.5        &  12.855\pmi 0.021 &11.932\pmi	 0.021 &11.404 \pmi	 0.020 &4 \\
28  &[RP93] 3       & 14.99   &K7    &  -27.0         &  11.204\pmi 0.022 &10.218\pmi	 0.021 & 9.582 \pmi	 0.020 &4 \\
29  &[RP93] 4       & 14.59   &K7-M0 &  -19.3         &  11.423\pmi 0.033 &10.669\pmi	 0.042 &10.244 \pmi	 0.031 &4 \\
30  &[RP93] 5       & 15.25   &K2-5  &  -126.9        &  11.959\pmi 0.026 &10.820\pmi	 0.030 &10.020 \pmi	 0.023 &4 \\
31  &[RP93] 6       & 14.21   &K7    &  -5.0          &  10.445\pmi 0.022 & 9.531\pmi	 0.022 & 9.111 \pmi	 0.025 &4 \\ 
32  &[RP93] 7       & 13.97   &K5    &  -9.8          &  11.491\pmi 0.023 &10.739\pmi	 0.023 &10.352 \pmi	 0.025 &4 \\ 
33  &[RP93] 8       & 15.33   &M1-2  &  -42.1         &  11.830\pmi 0.027 &11.066\pmi	 0.024 &10.832 \pmi	 0.021 &4 \\
	   	       
\hline	
\multicolumn{9}{|l|}{Notes:$^a$ 1: Hillenbrand \etal 1992, 2: Kim \etal 2005, 3: Petterson \etal 1987, 4:Reipurth \etal 1993, 5:Manoj \etal 2006} \\	 \multicolumn{9}{|l|}{JHK measurements from the 2MASS ALL-Sky Release Point Source Catalog.}
\end{tabular}
\end{minipage}
\end{table*}


There  are two YSOs,  NX Pup  and PHa  92, which  are of  some special
interest.    From    the   Figure   \ref{cg-1-plot.ps}    and   Figure
\ref{cg-22-plot.ps} it seems very likely that they are embedded in the
heads of CG 1 and CG  22 respectively.  Both the stars show associated
reflection nebulosities, \Ha emission  line and near IR excess.  Based
on these observations  Reipurth (1983) concluded that NX Pup  is an YSO
formed in  CG 1.  Sahu  \etal (1992) also  confirmed that PHa 92  is a
T-Tauri star formed  in CG 22. As these two  YSOs are still associated
with the respective CGs we can  use the proper motion of the two stars
as the  transverse motion  of the respective  CGs.  The  formation and
evolutionary models of the CGs predict a net radially outward motion .
The CG tails  are also directed radially outward.   It should be noted
that the tail  formation is a relatively fast  process and, therefore,
the tail direction is determined  by the current position of the cloud
and the exciting star, while the  direction of motion of the star born
in the  CG would be  determined by the  initial velocity of  the cloud
(inertia)  and  prolonged acceleration,  if  any,  due  to the  winds,
radiation and  supernova shocks of  massive stars.  Here we  have used
the proper motion measurements of  the YSOs to study the kinematics of
the CGs and effects of earlier  events in the region on these objects,
if any.


\section {Proper Motion of the Young Stellar Objects}

We have  collected the available  proper motion data on  the confirmed
YSO  candidates  from the    Naval  Observatory Merged  Astrometric
Dataset (NOMAD) (Zacharias \etal,  2005) catalog.  We have selected only
those YSOs  whose proper motion of  at least one  component is greater
than the error of the measurements given in the catalog.  We have also
considered the  catalog by Ducourant  (2005) for the proper  motion of
the Pre  Main Sequence (PMS)  stars.  The best  measurements (smallest
error)  available in the  two catalogs  have been  selected.  However,
(NOMAD) does not  have the measurements for NX Pup.  We have taken the
proper  motion measurements  for  NX  Pup from  Tycho  2 catalog  (Hog
\etal,2000). For the stars taken from NOMAD catalog, we have converted
their  proper motion  to the  galactic coordinates  by the  formula as
described by Mdzinarishvili (2005).

We tabulate the identification  number, name, radial distance from the
respective  star  to the  NOMAD  counterpart (r), equatorial  coordinates,
proper motion measurements and the associated errors in the respective
columns of Table \ref{proper_motion_data} .  The observed differential
proper motions  of the YSOs with  respect to the mean  proper motion of
the Vela OB2  are tabulated in Column (10) and  Column (11).  The proper
motion  measurements  in   galactic coordinates  are  included  in
Column (12) and Column (13) and  the differential proper motion of the
YSOs  with respect  to  the mean  proper  motion of  the  Vela OB2  in
galactic coordinate are tabulated in Column (14) and Column (15).

\begin{table}
\centering
\begin{minipage}{\linewidth}
\caption{Statistics of  proper motion of the YSOs and Vela OB2 members}\label{mean_pm_data}
\begin{tabular}{@{}lcccc@{}}
\hline
Objects & $<\mu_{\alpha}$ cos$\delta>$ & $\sigma(\mu_{\alpha} cos\delta)$ & $<\mu_{\delta}>$& $\sigma(\mu_{\delta})$ \\
        & mas yr$^{-1}$  &mas yr$^{-1}$          &mas yr$^{-1}$  & mas yr$^{-1}$  \\
\hline

All YSOS         &  -8.39 & 5.12     &  10.25  &   7.53        \\
High Velocity YSOs& -16.33& 6.47     &  21.65  &   8.23\\
Normal Velocity YSOs      &  -6.27 & 1.38     &   7.20   &   3.37        \\
Vela OB2 members &  -6.42 & 2.45     &   8.13  &   1.93        \\
\hline	
\end{tabular}
\end{minipage}
\end{table}

\begin{figure*}
\resizebox{16cm}{16cm}{\includegraphics{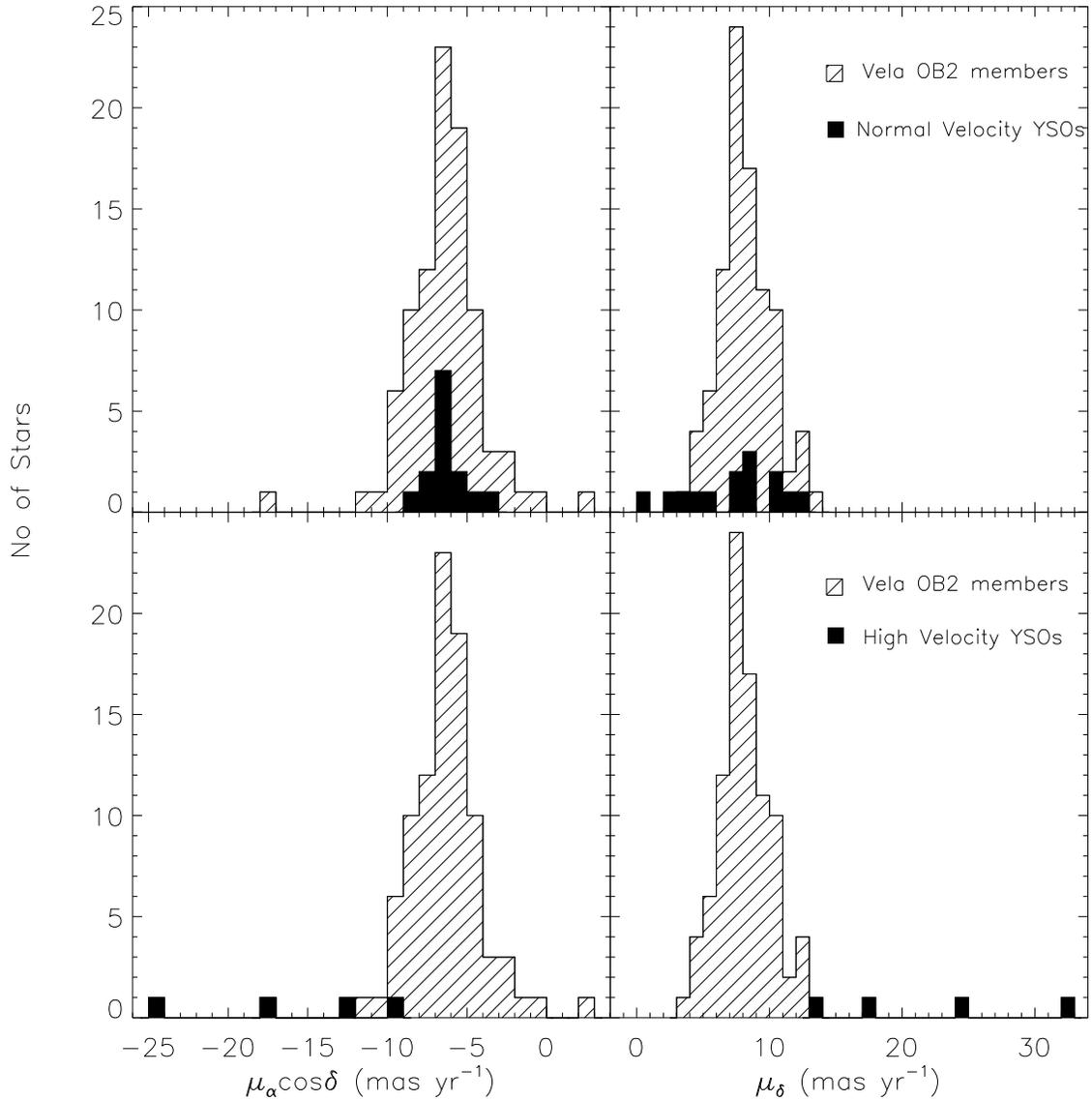}}
\caption{Proper motion histograms of \textit{Normal Velocity YSOs}, \textit{High Velocity YSOs},
and the members of Vela OB2 association. \textit{Line-filled} histogram represent the members of the association and the \textit{solid} histogram represent the YSOs.}
\label{hist-plot.ps}
\end{figure*}

We  have tabulated  the mean  proper motions  of the  YSOs  from Table
\ref{proper_motion_data} and  the member  of the Vela  OB2 association
(de Zeewu  \etal, 1999) from Tycho  2 Catalog (Hog,  2000). From Table
\ref{mean_pm_data} its quite clear that  mean proper motion of all the
YSOs is not similar to that  of the Vela OB2 members. We consider YSOs
2, 4,  11 and 29  as \textit{High Velocity  YSOs} due to  their proper
motions  being  greater than  2$\sigma$  from  the  mean in  both  the
components.  We discuss the properties of \textit{ High Velocity YSOs}
in detail in the Section 5.  Excluding the \textit{High Veocity YSOs}, the rest
of the  YSOs are  termed as \textit{Normal  Velocity YSOs}.   The mean
proper motion of  the \textit{Normal Velocity YSOs} is  similar to the
mean  motion of the  members of  the Vela  OB2 association  within the
error limit associated with the measurements of the proper motion.  We
have plotted the histogram of the proper motion of both \textit{Normal
  Velocity  YSOs} and \textit{High  Velocity YSOs}  with the  Vela OB2
association in Figure \ref{hist-plot.ps}.

The  heliocentric velocity  of  the association  V$_{helio}$  = 18  km
s$^{-1}$ (de Zeeuw\etal,  1999), which is equivalent to  V$_{LSR}$ = 4
km s$^{-1}$.  The  radial velocities of the CGs  and Diffuse Molecular
Clouds (DMCs) are taken from  Sridharan (1992), Woermann \etal ( 2001)
and   Otrupcek  \etal   (2000)  respectively.    The   mean  V$_{LSR}$
($<V_{LSR}>$) of the CGs is 1.74 km s$^{-1}$ with $\sigma_{V_{LSR}}$ =
4.73.  Among  all the CGs,  only CG 24  has radial velocity  more than
2$\sigma$ away from  the mean.  Excluding CG 24,  $<V_{LSR}>$ for rest
of the CGs is 2.13  with $\sigma_{V_{LSR}}$= 4.13.  $<V_{LSR}>$ of the
DMCs is 5 km s$^{-1}$ with $\sigma_{V_{LSR}}$ = 6 km s$^{-1}$.  Out of
the 106, 96 DMCs have  values within 2$\sigma$ of the calculated mean.
$<V_{LSR}>$   of   these   96   DMCs   is   3.5   km   s$^{-1}$   with
$\sigma_{V_{LSR}}$=  4.52 km  s$^{-1}$ .   These values  indicate that
there is a good coupling between  the CGs, DMCs and the members of the
Vela OB2  associations.  All  these systems of  objects have  the same
average motion.

\section{Discussion}

\begin{figure*}
\resizebox{16cm}{10cm}{\includegraphics{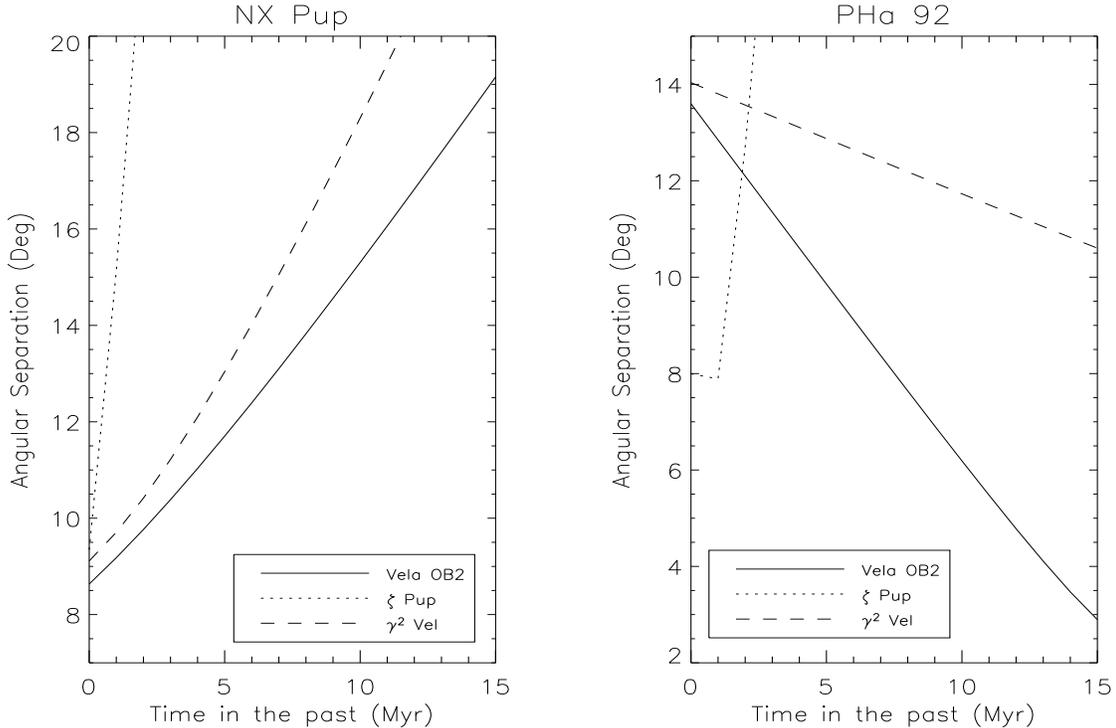}}
\caption{Angular separation  of NX  Pup and PHa  92 from  $\zeta$ Pup,
$\gamma^2$  Vel and  Vela OB2  association  in the  past. The  dotted,
dashed and  solid lines represent the  angular separation from
$\zeta$ Pup, $\gamma^2$ Vel and Vela OB2 association respectively. }
\label{angular-sep-l-b.ps}
\end{figure*}

Woermann \etal  (2001) have suggested that the  supernova explosion of
the massive binary companion of the $\zeta$ Pup is responsible for the
origin  of the  Gum Nebula.   They estimated  that the  explosion took
place \near 1.5  Myr ago. They also suggested  that the runaway O-star
$\zeta$ Pup was within $<$  0.5$^\circ$ of the expansion centre of the
neutral shell  \near 1.5 Myr  ago. As a  first order estimate  we have
traced back  the proper motion  of $\zeta$ Pup  and Pha 92 and  NX Pup
(embedded YSOs) in galactic  coordinates, and calculated their angular
separations. The angular separation of  PHa 92 and NX Pup from $\zeta$
Pup  as  a  function  of  time  in the  past  are  plotted  in  Figure
\ref{angular-sep-l-b.ps}.  We do not see any indication of convergence
between all three in the last  1.5 Myr .  This trend suggests that the
supernova explosion of  the companion of the $\zeta$  Pup did not give
any radial  motion to the  stars NX Pup  and PHa 92. So  the supernova
explosion was not responsible for the  bulk motion of CG 1 and CG 22
associated with NX Pup and PHa 92 respectively.

Woermann  \etal (2001)  also gave  a likely  configuration of  the Gum
nebula.   They  suggested  that most  of  the  CGs  and the  Vela  OB2
association are situated within the front face of the expanding shell.
The expected transverse velocity components  of the two stars based on
the radial  velocities of  the CGs and  the expansion velocity  of the
system can be calculated using the following formula
\hspace{7mm} $V_{T}=\sqrt{V^2_{exp}-V^2_{radial}} $.\\
\noindent  Considering  V$_{LSR}$  (CG  1)  =  +3.3  km  s$^{-1}$  and
V$_{LSR}$  (CG  22  )=  6.65  km  s$^{-1}$  from  Sridharan(1992)  and
$V_{exp}$  = 14 km  s$^{-1}$ for  the front  face from  Woermann \etal
(2001)  we obtained  the  expected transverse  velocity components  as
V$_{T}$  (CG  1)= 13.6  km  s$^{-1}$ and  V$_{T}$  (CG  22)= 12.32  km
s$^{-1}$. These values corresponds to  5.78 and 6.38 mas yr$^{-1}$ for
CG 1 and  CG 22 respectively with respect to  the centre of explosion.
Again, if CG 22 and CG 1 were close to the site of supernova explosion
of the companion of $\zeta$ Pup 1.5 Myr ago then the proper motions of
NX Pup and Pha 92 born in the respective CGs should be consistent with
their  motion  from  the  site  of  the  explosion  to  their  present
position. This would require a differential proper motion of NX Pup as
$\Delta{\mu_{l}cosb}$\hspace{1mm}\near
\hspace{1mm}-\hspace{1mm}30.09        mas       yr$^{-1}$,\hspace{3mm}
$\Delta{\mu_{b}}$\hspace{1mm}\near\hspace{1mm}-\hspace{1mm}36.23 mas yr$^{-1}$
and PHa 92 as $\Delta{\mu_{l}cosb}$\hspace{1mm}\near
\hspace{1mm}-\hspace{1mm}39.73        mas       yr$^{-1}$,\hspace{3mm}
$\Delta{\mu_{b}}$\hspace{1mm}\near\hspace{1mm}1.01 mas  yr$^{-1}$.  At present
we do  not have the proper  motion of the binary  companion of $\zeta$
Pup which  is the suggested centre  of the expansion. But  the other O
type  star $\gamma^2$  Vel shares  a similar  kind of  motion  as Vela
OB2. So it  is reasonable to adopt the mean proper  motion of Vela OB2
for the motion  of the centre.  Considering the  mean proper motion of
the   Vela  OB2,   the   expected   proper  motion   of   NX  Pup   is
$\mu_{l}cosb$\hspace{1mm}\near
\hspace{1mm}-\hspace{1mm}40.49        mas       yr$^{-1}$,\hspace{3mm}
$\mu_{b}$\hspace{1mm}\near\hspace{1mm}-\hspace{1mm}41.03 mas yr$^{-1}$
and for PHa 92 it is $\mu_{l}cosb$\hspace{1mm}\near
\hspace{1mm}-\hspace{1mm}50.13        mas       yr$^{-1}$,\hspace{3mm}
$\mu_{b}$\hspace{1mm}\near\hspace{1mm}2.31 mas yr$^{-1}$.

Considering the  earliest stars Yamaguchi (1999) suggested  the age of
the association  as less than 10  Myr. We also traced  back the proper
motion of Pha 92 and NX Pup from the $\gamma^{2}$ Velorum and the mean
proper  motion of  the  Vela OB2  association  up to  15 Myr.   Figure
\ref{angular-sep-l-b.ps} also shows the  angular separations of NX Pup
and PHa 92 from $\gamma^2$ Vel  and Vela OB2 association.  There is no
indication of convergence of NX Pup  with any of the sources ( $\zeta$
Pup,  $\gamma^2$  Velorum and  Vela  OB2  )  plotted in  the  diagram.
However, the  angular separation between the  PHa 92 and  the Vela OB2
association was shorter in the past than the present separation.

Adopting the proper  motion of Vela OB2 for the  motion of the centre,
the observed differential  proper motion of the stars  with respect to
the  Vela  OB2  association  are  estimated  and  tabulated  in  Table
\ref{proper_motion_data}  (column 15 and  16).  There  is significant
mismatch between the expected and observed differential proper motions
of the  YSOs with respect to the  Vela OB association. For  NX Pup the
expected   differential   proper   motion  is   $\Delta\mu_{l}cosb$
\near     \hspace{.5mm}     -\hspace{.5mm}30.09     mas     yr$^{-1}$,
$\Delta\mu_{b}$\near    \hspace{.5mm}    -\hspace{.5mm}36.23    mas
yr$^{-1}$   while   the   observed   differential  proper   motion   is
$\Delta\mu_{l}cosb$   \near\hspace{.5mm}    3.32   mas   yr$^{-1}$,
$\Delta\mu_{b}$ \near\hspace{.5mm} 0.04  mas yr$^{-1}$.  For PHa 92
the  expected  differential  proper motion  is  $\Delta\mu_{l}cosb$
\near\hspace{.5mm}       -\hspace{.5mm}39.73       mas      yr$^{-1}$,
$\Delta\mu_{b}$ \near\hspace{.5mm} 1.01 mas yr$^{-1}$
while  the observed differential  proper motion  is $\Delta\mu_{l}cosb$
\near    \hspace{.5mm}   1.90   mas    yr$^{-1}$,   $\Delta\mu_{b}$
\near\hspace{.5mm}  2.01 mas yr$^{-1}$.  For  other YSOs,  the expected
differential       proper       motion      $\mid$$\Delta\mu$$\mid$
\near  \hspace{1mm} 30  mas  yr$^{-1}$ while  the observed  differential
proper  motion  is $\mid$$\Delta\mu$$\mid$  \near \hspace{1mm}  5  mas
yr$^{-1}$. 


\begin{figure*}
\resizebox{16cm}{16cm}{\includegraphics{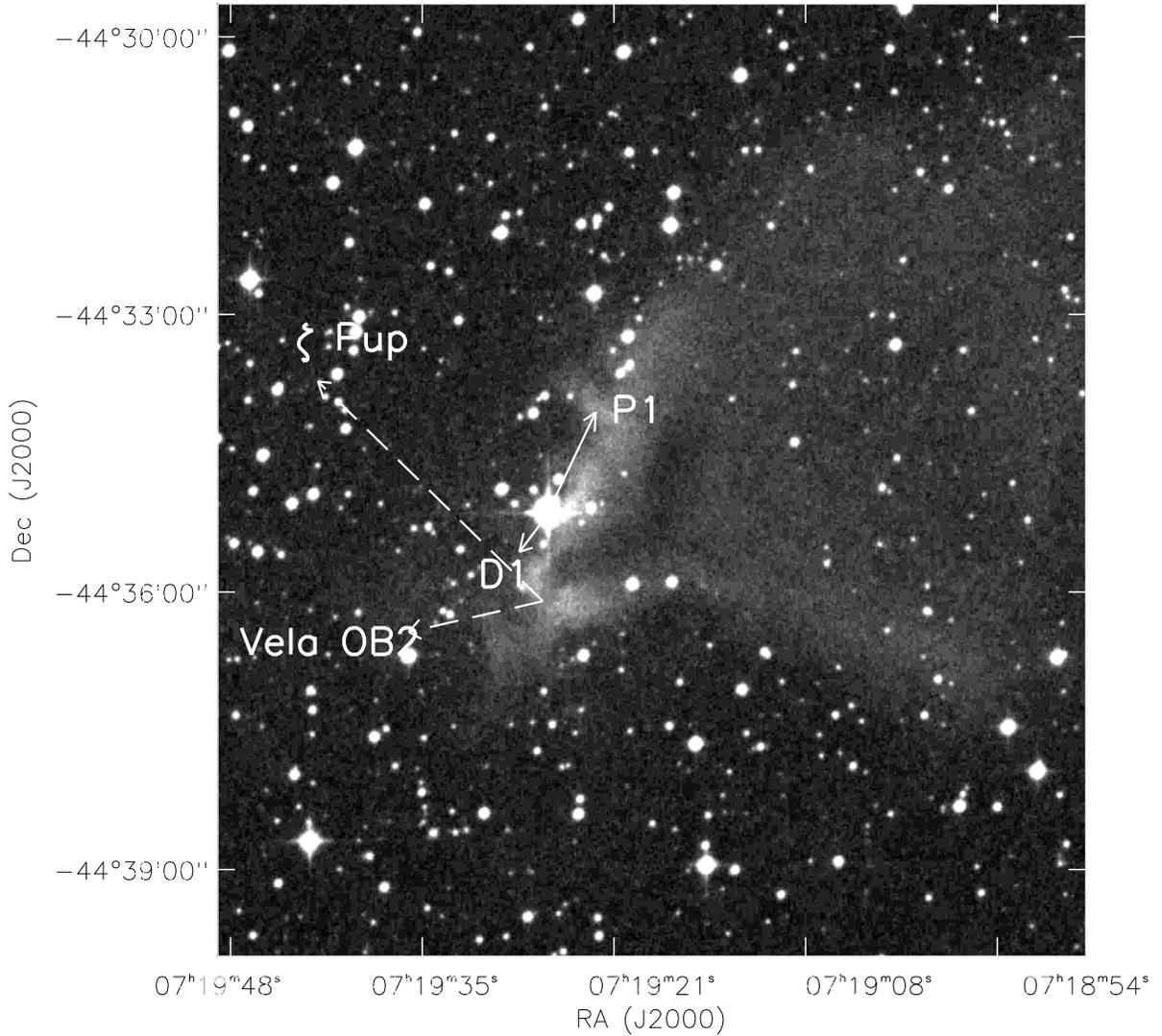}}
\caption{Proper motion vector \textit{P1} of NX Pup plotted on the DSS
  image of CG  1. The directions towards the  Vela OB2 association and
  $\zeta$  Pup   are  also   shown  by  \textit{dashed}   lines.   The
  differential proper motion vector \textit{D1} is also plotted. Arrow  lengths  are  proportional  to  the  respective  proper  motion. }
\label{cg-1-plot.ps}
\end{figure*}

\begin{figure*}
\resizebox{16cm}{16cm}{\includegraphics{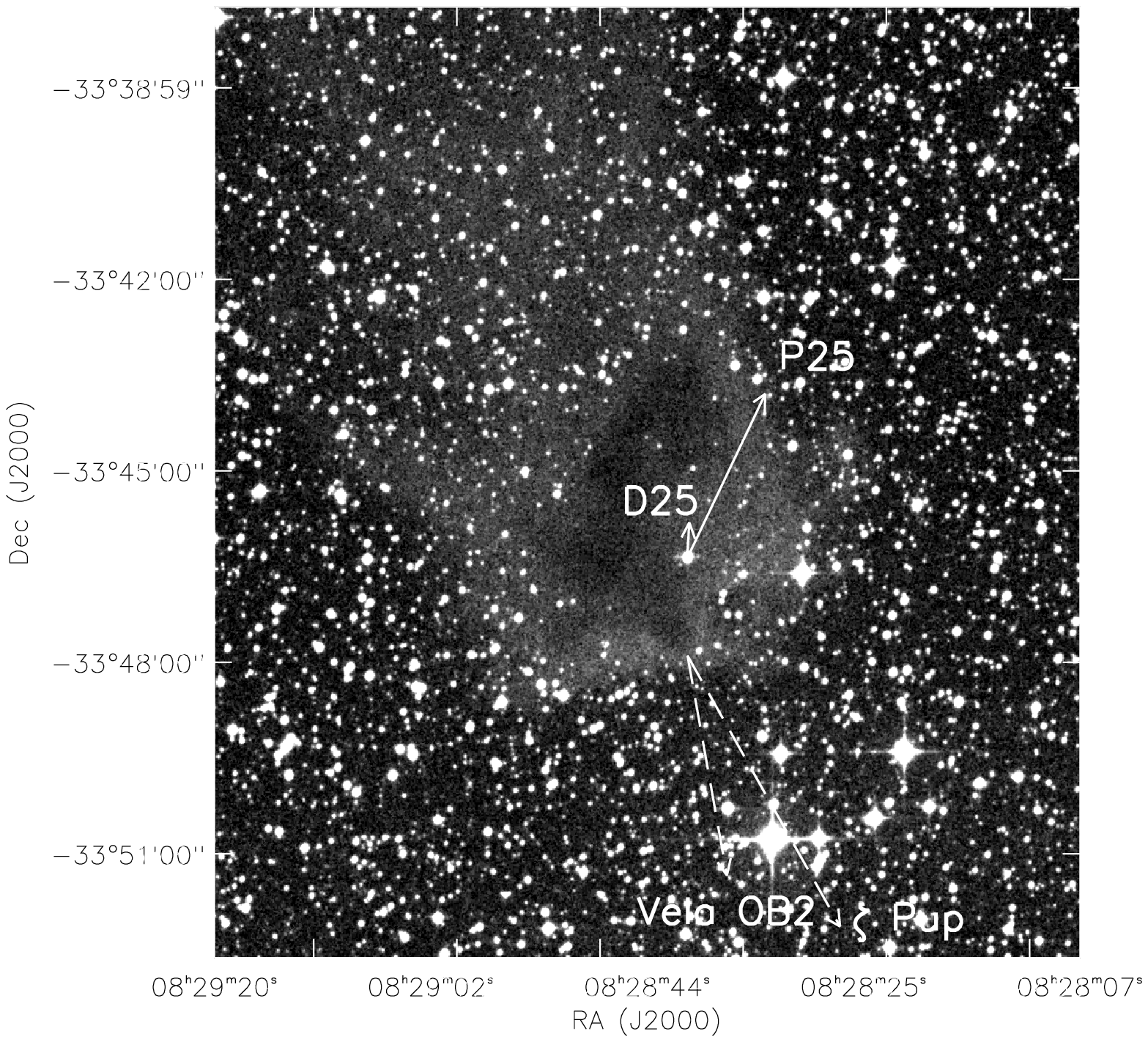}}
\caption{The proper motion vector \textit{P25} of PHa 92 plotted on
  the DSS  image of CG 22. The directions
  towards the Vela  OB2 association and $\zeta$ Pup  are also shown by
  \textit{dashed}  lines. The  differential proper  motion vector \textit{D25}
  is also plotted. Arrow  lengths are  proportional to  the
  respective proper motion. }
\label{cg-22-plot.ps}
\end{figure*}


The proper  motion vector \textit{P1}  of NX Pup  [Star identification
  No.  1 in Table \ref{spec_phot_data}] is plotted on the DSS image of
CG  1 in  Figure  \ref{cg-1-plot.ps}. The  differential proper  motion
vector \textit{D1} of  NX Pup with respect to  Vela OB2 association is
also  plotted  Figure \ref{cg-1-plot.ps}.   The  proper motion  vector
\textit{P25}  of  PHa  92   [Star  identification  No.   25  in  Table
  \ref{spec_phot_data}] is plotted on the DSS image of CG 22 in Figure
\ref{cg-22-plot.ps}.    The    differential   proper   motion   vector
\textit{D25} of  PHa 92  with respect to  the Vela OB2  association is
also  plotted in Figure  \ref{cg-22-plot.ps}.  The  directions towards
the  Vela OB2 association  and $\zeta$  Pup are  also shown  in Figure
\ref{cg-1-plot.ps} and  Figure \ref{cg-22-plot.ps}.  Within  the errors
associated  with the  proper motion  measurements we  do not  find any
clear evidence for the transverse motion for these two stars away from
the site of supernova explosion  of the companion of $\zeta$ Pup.  For
the  objects which  are not  embedded in  any CGs  but supposed  to be
associated with the Gum Nebula,  we get similar results.  They also do
not  show any  indication of  expansion  within the  errors of  proper
motion measurements.  This  suggests that the system of  CGs share the
mean motion of Vela OB2 and  do not show any systematic radial motions
away from the centre.

\begin{figure*}
\resizebox{12cm}{11.5cm}{\includegraphics{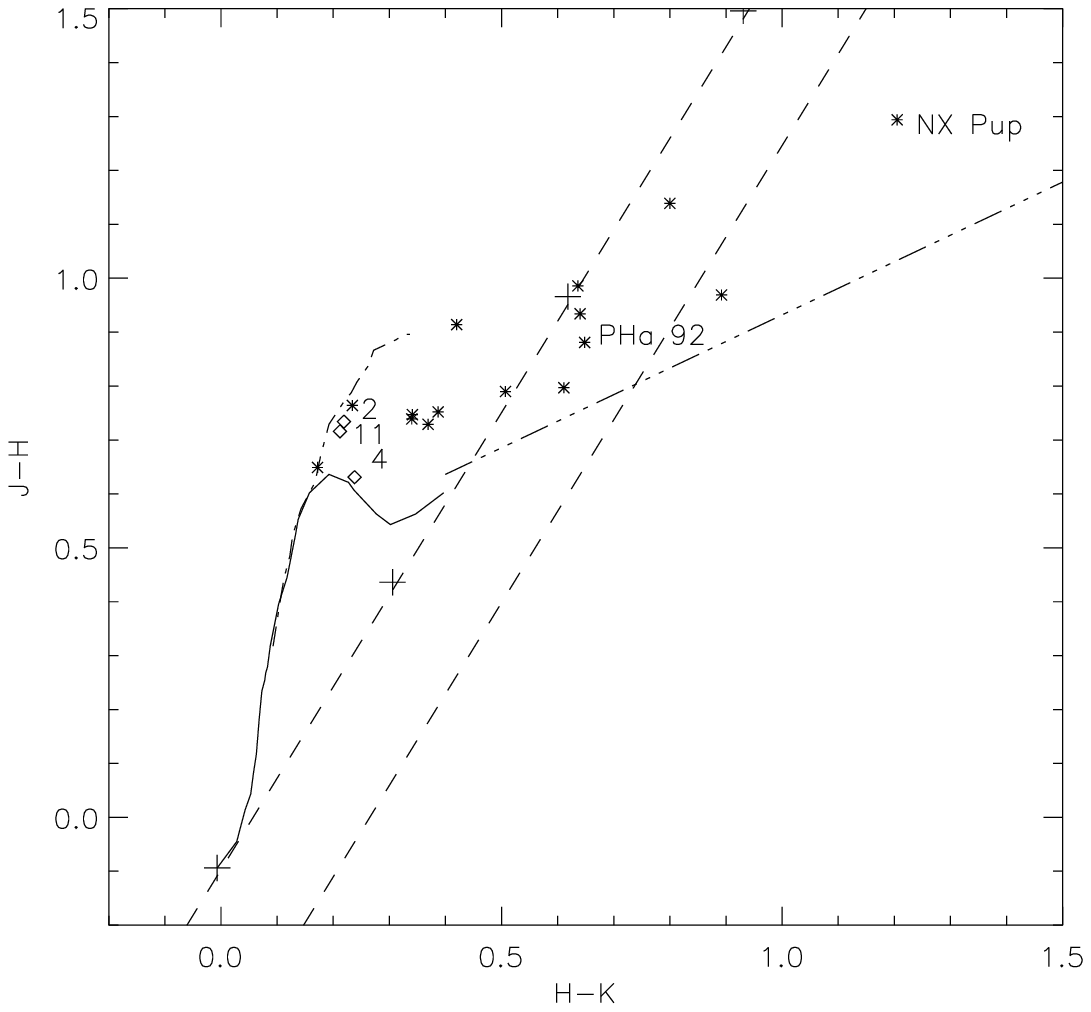}}
\caption{2MASS  JHK color-color diagram  for the  known YSOs  in and
  around the CGs with reliable NIR photometry. High velocity stars are
  marked  with  diamond  and  their corresponding  number  from  Table
  \ref{spec_phot_data}.   \textit{Solid} and \textit{dash  dot} curves
  are the locations  of main-sequence and giant stars  from Bessell \&
  Brett  (1988)  converted  into   2MASS  as  suggested  by  Carpenter
  (2001).  The  two \textit{dashed}  parallel  lines,  with the  slope
  derived from  interstellar reddening  law (Rieke \&  Lebofsky 1985),
  separate  CTTSs from  Herbig Ae/Be  and from  reddened main-sequence
  stars (Lee  \etal,  2000).  The  \textit{dash  dot dot}  line  is  the
  dereddened CTTS locus (Meyer \etal, 1997). Points marked with plus on
  the dashed line are at an interval of A$_{v}$ = 5 mag.  }
\label{2mass-ccd-spec-ysos.ps}
\end{figure*}

We have also found 3 high velocity  YSOs in CG 30 complex and one more
in Sa  101. They have  proper motion measurements more  than 3$\sigma$
away from the  the mean proper motion of the Vela  OB2.  For the stars
in the CG 30 complex Kim \etal (2005) have estimated an age of 2-5 Myr. These
YSOs  have H$\alpha$  emission  line  as well  as  Li absorption  line
indicating youth.  Figure \ref{2mass-ccd-spec-ysos.ps} gives a near IR
J-H vs.  H-K color-color (CC) diagram  of the known YSOs in and around
the   CGs   with  reliable   NIR   photometry   as   given  in   Table
\ref{spec_phot_data}.  High velocity stars are marked with diamond and
their  corresponding  identification  number,  as   given  in  the  Table
\ref{spec_phot_data}. We exclude the  high velocity star [RP93] 4 from
the  CC diagram  due to  poor  measurements.  The  location of  normal
main-sequence  and  giant  stars  from  Bessell \&  Brett  (1988)  are
modified to the 2MASS photometry  as suggested by Carpenter (2001) and
plotted with solid and dash dot lines respectively. Points marked with
$plus$ on the dashed line are at  an interval of A$_{v}$ = 5 mag.  Lee
\etal  (2005)  have  developed  an  empirical  and  effective  set  of
criteria, based on  the 2MASS colors, to select  candidate classical T
Tauri  stars (CTTSs).   They  found that  the  CTTS lie  approximately
between  the  two  parallel   dashed  lines,  defined  empirically  by
$(j\_m-h\_m)-1.7(h\_m-k\_m)+               0.0976=0$               and
$(j\_m-h\_m)-1.7(h\_m-k\_m)+0.450=0$, where $j\_m$, $h\_m$, and $k\_m$
are 2MASS magnitudes,  and the slope is specified  by the interstellar
reddening law (Rieke \& Lebofsky 1985).   The \textit {dash dot dot} line in the
CC diagram represents  the dereddened CTTS locus (Meyer  \etal 1997) ,
modified   to   the    2MASS   photometry   (Carpenter   ,2001)   with
$(j\_m-h\_m)-0.493(h\_m-k\_m)-0.439=0$.   The  dashed  parallel  lines
separate CTTS from Herbig Ae/Be and from reddened main sequence stars.
Positions  occupied by  the objects  in  the CC  diagram give  primary
information about the nature of the  YSOs.  NX Pup (Herbig Ae) and PHa
92 (T Tauri) both have near  IR excesses and their positions in the CC
diagram also match with the criteria suggested by Lee \etal (2005). It
is interesting to note that most of the YSOs which have similar proper
motion  as Vela  OB2, show  near IR  excesses but  all the  three high
velocity YSOs occupy  similar positions in the CC  diagram and they do
not  show  near-infrared  excesses.   But  there  are  no  significant
differences in H$\alpha$ or Li equivalent widths and radial velocities
of these objects compared with the other stars and the CGs (Kim \etal,
2005).  If they are indeed associated with the CG 30 complex then they
perhaps owe their high velocities  to the process of star formation in
the cloud  which is not understood. It  may also be possible that  the cause of
their high proper motion  affected their circumstellar environment and
as  a result  they do  not have  infrared excesses.   There  have been
suggestions  that during  the  star formation  in  binary or  multiple
systems YSOs  can gain high  velocities due to  dynamical interactions
[Sterzik \etal (1995), Gorti \& Bhatt (1996), Reipurth(2000)].  But CG
30  lies close  to  galactic plane  .   So it  is  also possible,  but
unlikely,  that   these  objects  are  foreground   YSOs  moving  with
relatively higher velocities and large proper motions.

\section{Conclusion}

In the  analysis presented  above no clear  evidence is found  for the
supernova  explosion of the  binary companion  of $\zeta$  Pup causing
expansion of the system of CGs in the Gum Nebula.  We also do not find
any systematic transverse expansion of the YSOs and the CGs in the Gum
Nebula.  It is possible that  the CGs retain the initial velocities of
their  parent  clumps  inside  the  GMC characterised  by  a  velocity
dispersion as seen in the radial velocities. The energy sources in the
Gum Nebula (stellar wind,  radiation and supernova explosions) perhaps
sweep  out the  diffuse  material  but not  the  relatively dense  and
massive CGs. The absence of CGs within some radius (\near\hspace{.6mm}
9 $^{\circ}$  ) of the  OB association would then  require destruction
due to  evaporation by the  UV radiation (Reipurth, 1983)  from the
central  energy sources.  No  clear evidence  is found  for transverse
motion of YSOs and CGs as predicted by RDI models.


\section{Acknowledgement}

This research  has made use of  the SIMBAD database,  operated at CDS,
Strasbourg, France.

\noindent This  publication makes  use of data  products from  the Two
Micron All Sky  Survey, which is a joint project  of the University of
Massachusetts    and   the    Infrared    Processing   and    Analysis
Center/California  Institute  of Technology,  funded  by the  National
Aeronautics  and   Space  Administration  and   the  National  Science
Foundation.

\noindent  The  Digitized  Sky  Surveys  were produced  at  the  Space
Telescope   Science  Institute  under   U.S.   Government   grant  NAG
W-2166. The  images of  these surveys are  based on  photographic data
obtained using  the Oschin Schmidt  Telescope on Palomar  Mountain and
the UK Schmidt Telescope.  The  plates were processed into the present
compressed digital form with the permission of these institutions.

\noindent  We   thank  the  referee, Prof.  W.  Zealey, for  critical
comments and valuable suggestions. 


\section*{REFERENCES}
Bertoldi, F. M., 1989, ApJ, 346, 735 \\
Bertoldi, F. M., \& McKee, C., 1990, ApJ, 354, 529\\
Bessell, M. S., Brett, J. M.1988, PASP, 100, 1134B\\
Brand, P. W. J. L., Hawarden, T. G., Longmore, A. J., Williams, P. M., Caldwell, J. A. R.,1983, MNRAS, 203, 215B \\
Carpenter,  J., M.,  2001, AJ, 121, 2851 \\
Chanot,  A., Sivan,  J. P., 1983, A\&A, 121, 19\\
Cutri,  R. M.  et al., 2003,  NASA/IPAC Infrared Science Archive  , The IRSA 2MASS All-Sky Point Source Catalog of Point Sources \\
Ducourant,  C., Teixeira,  R., Perie,  J. P., et al. 2005, A\&A, 438, 769 \\
de Zeeuw, P.T., Hoogerwerf,  R., de Bruijne,  J. H. J., Brown,  A. G. A., Blaauw,  A., 1999, AJ, 117, 354 \\
Elmergreen,  B. G.,1976, ApJ, 205, 405E \\
Gorti,  U., Bhatt, H. C., 1996, MNRAS, 278, 611G\\
Gum, C. S., 1952, Observatory, 72, 151\\
Hillenbrand,  L. A., Strom,  S. E., Vrba,  F. J., Keene,  J., 1992, ApJ, 397, 613 \\
Hog,  E.  et al., 2000, A\&A, 355, L27 
Kim,  J. S., Walter, F M., Wolk, S J., 2005, AJ, 129, 1564\\
Lee, Hsu-Tai, Chen, W. P., Zhang, Zhi-Wei, Hu, Jing-Yao, 2005, ApJ,624, 808L\\
Lefloch,  B.,  Lazareff, B.,  1994, A\&A, 289, 559\\
Manoj, P., Bhatt, H. C., Maheswar, G., Muneer, S.,2006, ApJ, 653, 657M\\
Miao, J., White, G. J., Nelson, R., Thompson, M., \& Morgan,  L., 2006, MNRAS, 369, 143\\
Mdzinarishvili,  T. G., Chargeishvili,  K. B., 2005, A\&A, 431, L1\\
Oort, J. H., \& Sitzer, L. Jr.,1955, ApJ,121,6\\
Otrupcek, R. E., Hartley, M., Wang, J.-S, 2000,PASA, 17, 92\\
Pozzo,  M., Jeffries,  R. D., Naylor,  T., Totten,  E. J., Harmer,  S., Kenyon,  M.  , 2000, MNRAS, 313, L23\\
Petterson, B., 1987, A\&A, 171, 101\\
Reipurth,  B., 1983, A\&A, 117, 183\\
Reipurth,  B., Petterson, B., 1993,  A\&A, 267, 439\\
Reipurth,  B., 2000, AJ, 120, 3177R \\
Rieke, G. H., Lebofsky, M.J., 1985, ApJ, 288,618 \\
Sahu, M., Sahu, K. C., 1992, A\&A, 259, 265S \\
Sridharan, T. K., 1992, JA\&A, 13, 217 \\
Sterzik, M. F., Durisen, R. H., 1995, A\&A, 304L, 9S \\
van der Hucht  et al., 1997, New Astron., 2, 245\\
Yamaguchi, N., Mizuno,  N., Moriguchi,  Y., Yonekura,  Y., Mizuno,  A., Fukui,  Y.  , 1999, PASJ, 51, 765\\
Zacharias,  N., Monet,  D., Levine,  S., Urban,  S., Gaume,  R., Wycoff,  G., 2004, BAAS, 36, 1418\\
Zealey, W. J., Ninkov, Z., Rice, E., Hartley, M., Tritton, S.B., 1983, Astrophys. Letters, 23, L119 

\clearpage
\pagestyle{empty}
\begin{landscape}
\center
\setcounter{table}{2}

\begin{table*}
\center
\caption{Star forming CGs and Diffuse Molecular Clouds}\label{cloud_param}
\begin{tabular}{@{}lccccccccc@{}}
\hline

CGs &  & & &\multicolumn{2}{|c|}{Vela OB2}&\multicolumn{2}{|c|}{$\gamma^{2}$ Vel}& \multicolumn{2}{|c|}{$\zeta$ Pup}\\
and  &   l        & b        &  $V_{LSR}$ &  Angular     &  Projected  &   Angular  &    Projected &   Angular   &  Projected \\
the  &            &          &            &  separation &   distance   & separation &     distance & separation  &  distance\\
DMCs &($^\circ$)& ($^\circ$) & $km$$ s^{-1}$&  ($^\circ$) &   $(pc)$      &($^\circ$) &   $(pc) $      & ($^\circ$)    &   $(pc)$\\
\hline
CG 1             & 256.14 & -14.07 & 3.3     &8.6   & 67   &9.11     &71   &9.35  &73 \\
CG 30/31/38      & 253.29 & -1.61  & 5.8-7   &10.67 & 84   &11.3     &89   &4.11  &32 \\
CG 22            & 253.58 & +2.96  & 6.5-6.8 &13.54 & 106  &14.1     &110  &7.97  &63 \\
CG 13            & 259.48 & -16.43 & 3.7     &9.03  &71	   &9.29     &73   &12.22 &96 \\
CG4/CG6/Sa101    & 259.48 & -12.73 & 0.9-1.7 &5.98  &47	   &6.38     &50   &9.03  &71\\
\hline
\multicolumn{10}{|l|}{Note: $V_{LSR}$ from Sridharan (1992). Projected distances are at 450 $pc$ }.
\end{tabular}
\end{table*}

\begin{table*}
\center
\caption{Proper Motion  of the YSOs associated with Cometary Globules and Diffuse Molecular Clouds}\label{proper_motion_data}
\begin{tabular}{@{}lllllllllllllll@{}}
\hline
No. &Name       &r       & RA  & Dec  &$\mu_{\alpha} cos\delta$ &  e  &$\mu_{\delta}$   &e   & $\Delta\mu_{\alpha} cos\delta$  &$\Delta\mu_{\delta}$  & $\mu_{l} cosb $  & $\mu_{b}$ & $\Delta\mu_{l} cosb$ & $\Delta\mu_{b}$  \\
    &           &$arcsec$&$(2000)$ &$(2000)$ &$ mas\hspace{1mm} yr^{-1}$&$ mas\hspace{1mm} yr^{-1}$ &$ mas\hspace{1mm} yr^{-1}$ &$ mas\hspace{1mm} yr^{-1}$ &$ mas \hspace{1mm}yr^{-1}$ &$ mas\hspace{1mm} yr^{-1}$&$ mas\hspace{1mm} yr^{-1}$& $ mas\hspace{1mm} yr^{-1}$&$ mas \hspace{1mm}yr^{-1}$&$ mas\hspace{1mm} yr^{-1}$ \\
\hline

1  &NX Pup     & 0.123  & 07 19 28.26& -44 35 11.4 &   -4.2    &     2.0  &       6.1  &       1.9  &   2.22&  -2.03  &     -7.08  &     -1.26 & 3.32&    0.04\\
3  &KWW 1892   & 0.101  & 08 08 22.15& -36 03 47.0 &   -6.5    &     4.8  &       7.7  &       4.7  &  -0.08&  -0.43  &	    -9.99  &     -1.29 & 0.41&    0.01\\
6  &KWW 1637   & 0.259  & 08 08 39.27& -36 05 01.7 &   -7.7    &     2.8  &       8.6  &       2.7  &  -1.28&   0.47  &	   -11.40  &     -1.81 & 1.00&   -0.51\\
7  &KWW 873    & 0.024  & 08 08 45.40& -36 08 40.2 &   -8.4    &     4.7  &       4.6  &       4.7  &  -1.98&  -3.53  &	    -8.42  &     -4.56 & 1.98&   -3.26\\
8  &KWW 1043   & 0.354  & 08 08 46.82& -36 07 52.8 &   -7.0    &     4.7  &       8.5  &       4.8  &  -0.58&   0.37  &	   -10.94  &     -1.27 & 0.54&    0.03\\
9  &KWW 975    & 0.099  & 08 08 33.87& -36 08 09.8 &   -5.9    &     4.8  &      12.3  &       4.8  &   0.52&   4.17  &	   -13.53  &      1.71 & 3.13&    3.01\\
20 &PHa 51     & 0.069  & 08 12 47.04& -36 19 17.9 &   -6.4    &     4.8  &       0.5  &       4.6  &   0.02&  -7.63  &	    -3.98  &     -5.03 & 2.59&   -0.37\\
22 &PHa 34     & 0.061  & 08 13 56.07& -36 08 02.1 &   -5.7    &     4.7  &       5.6  &       4.8  &   0.72&  -2.53  &	    -7.81  &     -1.67 & 4.06&   -3.37\\
24 &PHa 41     & 0.014  & 08 15 55.32& -35 57 58.1 &   -7.4    &     4.6  &       2.7  &       4.6  &  -0.98&  -5.43  &	    -6.34  &     -4.67 & 6.42&   -3.73\\
25 &PHa 92     & 0.041  & 08 28 40.70& -33 46 22.3 &   -6.6    &     4.6  &      10.4  &       4.6  &  -0.18&   2.27  &	   -12.30  &      0.71 & 1.90&    2.01\\
28 &[RP93] 3   & 0.987  & 07 31 10.81& -47 00 32.5 &   -6.7    &     5.0  &      10.9  &       4.9  &  -0.28&   2.77  &	   -12.73  &     -1.29 & 2.33&    0.01\\
30 &[RP93] 5   & 1.104  & 07 31 36.68& -47 00 13.2 &   -6.9    &     5.0  &       8.4  &       5.0  &  -0.48&   0.27  &	   -10.57  &     -2.54 & 0.17&   -1.24\\
31 &[RP93] 6   & 0.541  & 07 31 37.42& -47 00 21.5 &   -7.0    &     4.9  &       3.6  &       5.0  &  -0.58&  -4.53  &	    -6.29  &     -4.73 & 4.11&   -3.43\\
32 &[RP93] 7   & 1.501  & 07 33 26.86& -46 48 42.6 &   -3.2    &     4.9  &      11.1  &       4.9  &   3.22&   2.97  &	   -11.37  &      2.03 & 0.97&    3.33\\
33 &[RP93] 8   & 0.363  & 07 15 40.89& -48 31 27.3 &   -4.5    &     5.8  &       7.1  &       5.9  &   1.92&  -1.03  &	    -8.29  &     -1.39 & 2.11&   -0.09\\
\textbf{2}  &\textbf{KWW 464} & 0.404 & 08 08 00.66 &  -35 57 33.3 & -17.4&       4.7  &      17.5  &  4.7  & -10.98&   9.37  &	   -24.13  &     -5.17 & -13.73&   -3.87\\
\textbf{4}  &\textbf{KWW 598} & 1.089 & 08 08 37.60 &  -36 09 49.4 & -10.0&      18.0  &      24.0  &  5.0  &  -3.58&  15.87  &	   -25.59  &      4.61 & -15.19&    5.91\\
\textbf{11} &\textbf{KWW 1953}& 0.103 & 08 08 26.93 &  -36 03 35.4 & -13.0&       4.9  &      13.1  &  5.0  &  -6.58&   4.97  &	   -18.05  &     -3.83 & -7.65&   -2.53\\
\textbf{29} &\textbf{[RP93] 4}& 1.177 & 07 31 21.85 &  -46 57 43.9 & -24.9&       4.9  &      32.0  &  4.9  & -18.48&  23.87  &	   -39.66  &     -8.47 & -29.26&   -7.17\\

\hline
\multicolumn{15}{|l|}{Note: Differential proper motions are given with respect to Vela OB2 association. The \textit{High Velocity YSOs} (2, 4, 11, 29) have been highlighted.}
\end{tabular}
\end{table*}

\end{landscape}

\label{lastpage} 
\end{document}